\documentclass{article}
\usepackage{graphicx,amssymb,amsmath,amsfonts,bm}
\usepackage[latin1]{inputenc}
\usepackage[T1]{fontenc}
\usepackage{epsfig}
\usepackage{graphicx}
\usepackage{epsfig}
\usepackage{dcolumn}
\usepackage{theorem}
\usepackage{amsfonts}
\usepackage{float}
\usepackage{color}

\begin{document}
\title{Kinetic models for dilute solutions of dumbbells in non-homogeneous flows revisited}

\author{Pierre Degond\thanks{
Université de Toulouse, UPS, INSA, UT1, UTM, Institut de Mathématiques de Toulouse, F-31062, Toulouse, France \& CNRS, Institut de Mathématiques de Toulouse, UMR 5219, F-31062 Toulouse, France. e-mail: pierre.degond@math.univ-toulouse.fr
}, 
Alexei Lozinski\thanks{
Université de Toulouse, UPS, INSA, UT1, UTM, Institut de Mathématiques de Toulouse, F-31062, Toulouse, France \& CNRS, Institut de Mathématiques de Toulouse, UMR 5219, F-31062 Toulouse, France. e-mail: alexei.lozinski@math.univ-toulouse.fr
} 
and Robert G. Owens\thanks{
D{\'e}partement de math{\'e}matiques et de statistique, Universit{\'e} de Montr{\'e}al,
CP 6128 succ. Centre-Ville, Montr{\'e}al QC H3C 3J7, Canada. e-mail: owens@dms.umontreal.ca
}
}
\date{}

\maketitle

\begin{abstract}
We propose a two fluid theory  to model a dilute polymer solution assuming that it consists of two phases, polymer and solvent, with two distinct macroscopic velocities. The solvent phase velocity is governed by the macroscopic Navier-Stokes equations with the addition of a force term describing the interaction between the two phases. The polymer phase is described on the mesoscopic level using a dumbbell model and its macroscopic velocity is obtained through averaging. We start by writing down the full phase-space distribution function for the dumbbells and then obtain the inertialess limits for the Fokker-Planck equation and for the averaged friction force acting between the phases from a rigorous asymptotic analysis. The resulting equations are relevant to the modelling of strongly non-homogeneous flows, while the standard kinetic model is recovered in the locally homogeneous case.
\end{abstract}

\section{Introduction}
The importance and continued research interest in the non-homogeneous flows of suspensions and polymeric fluids may be attributed to the richness of the fluid mechanical phenomena often associated with such flows (such as migration, slip and the F{\aa}hraeus-Lindqvist effect \cite{FL31}) as well as their prevalence in polymer processing, oil recovery, biorheology and microfluidic devices, for example. In this paper we shall be concerned with the rigorous mathematical derivation of the governing equations for non-homogeneous flows of dilute solutions of FENE-type dumbbells.

There is some agreement on the macroscopic equations describing the rheology of non-homogeneous flows of dilute polymer solutions. The issue of the compatibility of three different formulations of these equations was addressed by Beris and Mavrantzas \cite{BerisMav94} in 1994. The governing equations, specifically those for the elastic stress tensor and the polymer number density, were obtained in the references cited by the authors using an non-homogeneous kinetic theory \cite{Bhaveetal91}, a continuum two-fluid Hamiltonian model \cite{MavBeris92} and a body tensor formulation in continuum mechanics \cite{Ottinger92}. Beris and Mavrantzas showed that a stress diffusion term that appeared in the constitutive equations derived by Bhave et al. \cite{Bhaveetal91} and Öttinger \cite{Ottinger92} also arose in their two-fluid Hamiltonian model, provided that a particular dissipation term was kept in their derivation and not omitted as in the original paper \cite{MavBeris92}. A term that was quadratic in the stress appeared in the constitutive equation of Öttinger \cite{Ottinger92} but not in the other two formulations. This term, however, would not lead to qualitative changes in the predictions of the three models, at least for small values of the velocity gradient. We note here that the derivation using a kinetic theory model of a closed-form constitutive equation for Hookean dumbbells by Bhave et al. \cite{Bhaveetal91} was only possible because of their use of the homogeneous form of the Kramers expression (Eqn. (40) of their paper). After redeveloping and correcting the analysis of Bhave et al. \cite{Bhaveetal91}, Beris and Mavrantzas showed that the polymer number density equation was the same for all three formulations.

An important phenomenon associated with nonhomogeneous flows of polymer solutions is that of \textit{molecular migration}. By this we mean that the centre of mass of a molecule does not move at all times with the same velocity as that of the surrounding solvent at the position of the centre of mass. The argument of Tirrell and Malone \cite{TirrellMalone77}, among others, that a macromolecule will try to minimize its free energy by moving to a region where stresses are lower was rejected by Aubert and Tirell \cite{AubertTirrell80} and Aubert et al. \cite{Aubertetal80} who concluded that there was no mechanism for coupling the free energy driving force to the motion of the centre of mass. Aubert and Tirrell \cite{AubertTirrell80} showed that for a free-draining dilute solution of bead-spring chains where the solvent velocity $\bm{v}_s$ could be expressed as a second order Maclaurin series
\[\bm{v}_s(\bm{x},t) = \bm{v}(\bm{0},t) + \bm{x}\cdot\nabla\bm{v}_s(\bm{0},t) + \frac{1}{2}\bm{xx}:\nabla\nabla\bm{v}(\bm{0},t),\]
the only migration of molecules could be in directions where $\bm{v}_s$ has non-zero Cartesian components. In particular, cross-streamline migration was stated to never occur in rectilinear flows (i.e. parallel flows with only one Cartesian component) but could occur in curvilinear flows. The authors then illustrated their point by showing that in Poiseuille flow of a dilute polymer solution no migration occurred in the radial direction, although the polymer lagged behind the solvent along streamlines. In circular Couette flow, however, it was demonstrated that cross-stream migration took place. The analysis of Aubert and Tirrell \cite{AubertTirrell80} was generalised, with similar conclusions, by Sekhon et al. \cite{Sekhonetal82} and Brunn and Chi \cite{BrunnChi84}. These authors showed, however, that the introduction of hydrodynamic interaction made cross-stream migration possible even in parallel flows. A similar point was made more recently by Ma and Graham \cite{MaGraham05} who stated that in the absence of hydrodynamic interactions between polymer segments no migration is found in shear flows without streamline curvature. It should be noted here, however, that wall effects on the cross-stream gradient of polymer stress were not considered in arriving at these conclusions. In the work of Beris and Mavrantzas \cite{BerisMav94}, Bhave et al. \cite{Bhaveetal91} and Öttinger \cite{Ottinger92}, discussed above, cross-stream migration was possible in wall-bounded non-homogeneous flows because of a term proportional to the divergence of the polymer conformations in the definition of the polymer mass flux, the conformation tensor being anisotropic due to the presence of the walls. In the Fokker-Planck-based approach of Lozinski et al. \cite{Lozinskietal04} to the problem of plane Poiseuille flow of a dilute polymer solution, similar considerations of wall effects on the allowable polymer conformations led to a non-homogeneous flow and cross-stream migration of polymer away from the walls. Stress-induced polymer migration in a Couette device has been studied by Apostolakis et al. \cite{ApostMavBeris02} using the two-fluid Hamiltonian model of Mavrantzas and Beris \cite{MavBeris92}. The authors found that at large Deborah numbers polymer chains migrated from the outer towards the inner cylinder, in qualitative agreement with the theoretical predictions of Aubert and Tirrell \cite{AubertTirrell80} and the pioneering experimental results of Shafter et al. \cite{Shafer74} and Dill and Zimm \cite{DillZimm79} showing DNA migration in high molecular weight DNA solutions between rotating concentric cylinders. In the present paper, the Kramers-type expression derived in Section \ref{ssec:mbeK} for the elastic stress takes into account the presence of walls and the consequent restrictions that this imposes on molecule configurations. Moving in a direction normal to the wall and into the fluid will generally mean that the configurations available to a dumbbell change and with this the evaluation of the stress tensor. It is the variation of stress in the normal direction to the wall that drives migration and leads to a non-uniform number density in the present case. An expression for the difference in the polymer and solvent velocities when the fluid surrounding a dumbbell is considered to be the solvent alone is given in Section \ref{ssec:Nde}. Although no explicit account is taken of hydrodynamic interaction in the developments of Sections \ref{sec:KT} and \ref{sec:iless}, we show in Section \ref{sec:polymandsolvvels} that considering the fluid surrounding a dumbbell to be the polymer solution leads to an expression for the perturbation to the solvent velocity that is proportional to the hydrodynamic force exerted by the dumbbells on the solution.

Although the present paper is in the first place concerned with the derivation of a Fokker-Planck equation for dumbbells of FENE type in a non-homogeneous flow when bead inertia may be neglected, the underlying governing equations with the full account of the bead inertia are described in detail in Section \ref{sec:KT}. The derivation of the equations in the inertialess limit is performed in Section \ref{sec:iless}. In Section \ref{ssec:FPe} we follow Degond and Liu \cite{DegondLiu09} in writing the full phase-space distribution function as a perturbation series about its inertialess value and show rigorously that the inertialess case is just the limit as the bead mass goes to zero of the full set of equations when inertia is included. The same result was obtained by Schieber and Öttinger \cite{JSO88} who also concluded that both equilibration in momentum space and neglect of the acceleration term in the bead equations of motion are consequences of allowing the bead mass to tend to zero, and not independent assumptions as stated by Bird et al. \cite{Birdetal87} and Curtiss et al. \cite{Curtissetal76}. The two fluid theory of Doi \cite{Doi90}, Doi and Onuki \cite{DoiOnuki92} and Milner \cite{Milner91} has been influential in our approach to deriving the equation of linear momentum and of a Kramers expression for the elastic stress tensor in Section \ref{ssec:mbeK}. The solution is viewed as two coexisting continuous media, the polymer and the solvent, moving with different velocities. A separate momentum balance equations is written down for the solvent phase while the macroscopic average velocity of the polymer phase is obtained by averaging from the mesoscopic description. In our derivation we mostly suppose that the fluid surrounding a dumbbell in dilute polymer solutions should be taken to be the solvent and not the volume fraction average of the solvent and polymer, in contrast with some other kinetic theory descriptions of dilute polymer solutions \cite{BerisMav94,Bhaveetal91}. We show, however, that assuming that the velocity of the surrounding fluid is the same as that of the solution as a whole is also possible in our framework, cf. Section \ref{sec:polymandsolvvels}. The restriction of conformations available to model polymers near a solid wall mean that our derived Kramers-type expression for the elastic stress tensor implies, as mentioned above, a stress gradient in the normal direction to a wall and further, that off diagonal components (in particular, the elastic shear stress) are zero there. Sufficiently far from the wall, or in homogeneous flows, our Kramers-type expression simplifies to the usual one found in the literature \cite{Birdetal87,Kramers44}.

In reworking the kinetic theory of Bhave et al. \cite{Bhaveetal91} for non-homogeneous flow of dilute polymer solutions, Beris and Mavrantzas \cite{BerisMav94}, referred to above, used truncated Taylor series about some spatial position vector $\bm{r}$ to obtain an expression for the polymer mass flux at $\bm{r}$ and then, using the polymer species conservation equation, wrote down an equation for the dumbbell number density which was common to all three formulations \cite{Bhaveetal91,MavBeris92,Ottinger92} considered. In Section \ref{ssec:Nde} we obtain the number density equation without resort to approximations and find that, if we chose the fluid velocity to be that of the solution rather than of the solvent, the exact number density equation would be precisely that derived in Beris and Mavrantzas's paper. This might explain why, despite differences in the constitutive equations appearing in \cite{Bhaveetal91,MavBeris92,Ottinger92}, there is perfect agreement on the number density equation.

In the homogeneous flow of a FENE-P or Oldroyd B fluid, substitution of the Kramers expression for the elastic stress tensor into the evolution equation for the dumbbell conformation tensor yields a closed-form constitutive equation. We seek to show that the kinetic theory presented here does not allow for such a closed-form constitutive equation in the case of non-homogeneous flows without resorting to closure or some other approximations, however. An alternative to a closure approximation would be to follow Bhave et al. \cite{Bhaveetal91} and simply employ the homogeneous form of the Kramers expression for the elastic stress tensor. This would, however, be inconsistent with retaining a diffusion term in the non-homogeneous Fokker-Planck equation, as we take pains to show in Section \ref{ssec:elaborateeqns}.

\section{Kinetic theory}\label{sec:KT}

\begin{figure}
\begin{center}
{\includegraphics{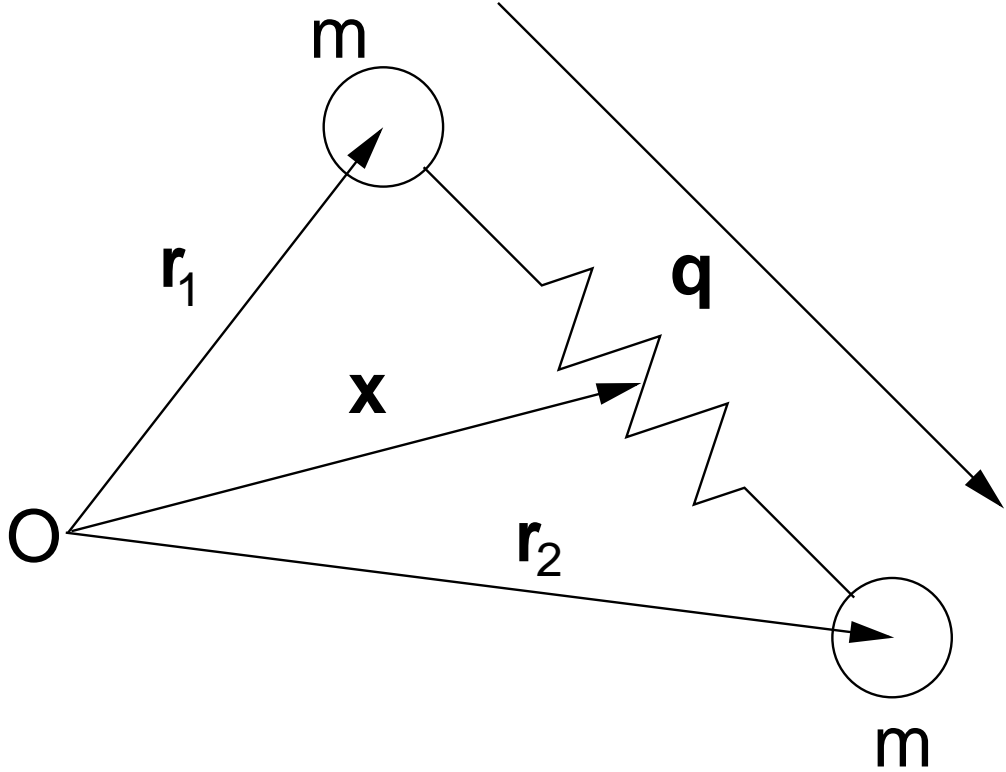}}
\end{center}
\caption{Elastic dumbbell consisting of two point masses of mass $m$, joined by a massless spring. $\bm{x}$ denotes the position vector of the centre of mass and $\bm{q}$ the end-to-end vector.}
\label{fig:dumbbell}
\end{figure}

We consider a dilute polymer solution filling a domain $\Omega\subset\mathbb{R}^d$. In the spirit of the two fluid theory of Doi \cite{Doi90}, Doi and Onuki \cite{DoiOnuki92} and Milner \cite{Milner91} we assume the solution to consist of two phases, polymer and solvent, with two distinct macroscopic velocity fields $\bm{v}_p$ and $\bm{v}_s$, respectively. The description of the solution is completed by specifying the polymer volume fraction $\varphi$ (and, consistently, the solvent volume fraction $1-\varphi$) which is allowed to vary in space and time.  The solvent phase is assumed to be an incompressible Newtonian fluid and its velocity is governed by the macroscopic Navier-Stokes equations with the addition of the force term describing the interaction between the two phases. In contrast, the polymer phase is modelled on the mesoscopic  level and the macroscopic quantities $\bm{v}_p$ and $\varphi$ are obtained through averaging. More specifically, the polymer molecules are represented by dumbbells consisting of two beads (more precisely, of two point masses), each of mass $m$, joined by a massless spring, as shown in Fig. \ref{fig:dumbbell}. The position vector of bead $i$ ($i=1,2$) is denoted by $\bm{r}_i(t)$ and its velocity by $\bm{V}_i =\dot{\bm{r}}_i$. The velocity of the surrounding fluid, i.e. that of the solvent, at the point having position vector $\bm{r}_i$ is denoted by $\bm{v}_i=\bm{v}_s(\bm{r}_i)$. The equations of motion of the $i$th bead may therefore be written down as a first-order system in the following way:
\begin{align}
m d\bm{V}_i & = \left(-\zeta(\bm{V}_i - \bm{v}_i) + \bm{F}_i\right)\;dt + \sqrt{2\zeta{k_B}T}d\bm{W}_i,\label{beadeqns1}\\
d\bm{r}_i & = \bm{V}_i\;dt,\label{beadeqns2}
\end{align}
where $\zeta$ denotes a drag coefficient, $\bm{F}_i$ is the spring force acting on bead $i$ and $\bm{W}_1$, $\bm{W}_2$ are two mutually independent $d$-dimensional Wiener processes. We assume that $\bm{F}_1=-\bm{F}_2=\bm{F}$, where $\bm{F}$ is a function of the dumbbell end-to-end vector $\bm{r}_2-\bm{r}_1$ alone.
$k_B$ and $T$ are, in the usual notation, the Boltzmann constant and the (absolute) temperature. 

% In writing down (\ref{beadeqns1})-(\ref{beadeqns2}) we ignore hydrodynamic interaction and assume that the drag on the $i$th bead is directly proportional to the velocity of the bead relative to that of the surrounding fluid.
\textbf{Remark 1}\label{ftnote}\textit{
The velocity $\bm{v}_i=\bm{v}_s(\bm{r}_i)$ in (\ref{beadeqns1}) should be understood as that of the solvent averaged over a box of size greater than that of a bead but smaller than the distance between the two ends of a dumbbell. We further assume that the fluctuations of the solvent velocity around $\bm{v}_s$  can be disregarded over such a box. This means, in particular, that the mechanism of hydrodynamic interaction is not included in our picture, i.e. the fluctuations of the solvent velocity induced by the vibrations of a bead of a dumbbell are assumed to fade away on a length smaller then the distance to the other bead. We admit that this simplifying assumption of separation of scales is not necessarily satisfied in reality. A more adequate modelling may be achieved in the approach of Coupled Langevin Equations \cite{OoFr81}, \cite{OtRa89} where the solvent velocity is treated in essentially the same way as the polymer one, i.e. its random fluctuations on the length scale of a dumbbell are not neglected so that $\bm{v}_s$ satisfies a stochastic (partial) differential equation. Such an approach is beyond the scope of the present paper and we content ourselves here with only a macroscopic description of the solvent, treating only its averaged velocity. The choice of $\bm{v}_s$ in the drag force in (\ref{beadeqns1}) is thus somewhat arbitrary. One could consider other options such as, for example, assuming that the averaged velocity of the surrounding fluid is that of the entire polymer-solvent solution. We reexamine  this option in Section \ref{sec:polymandsolvvels}. 
}

We now introduce the $2d$-dimensional vectors ${\bm{r}}=(\bm{r}_1^T,\bm{r}_2^T)^T$, ${\bm{V}}=(\bm{V}_1^T,\bm{V}_2^T)^T$, ${\bm{F}}=(\bm{F}_1^T,\bm{F}_2^T)^T$, ${\bm{v}}=(\bm{v}_1^T,\bm{v}_2^T)^T$ and denote by $\Psi({\bm{r}},{\bm{V}},t)$ the dumbbell distribution function, defined such that $\Psi({\bm{r}},{\bm{V}},t)\;d{\bm{r}}\;d{\bm{V}}$ is the expected number of dumbbells having bead position vectors in the interval $[{\bm{r}},{\bm{r}}+d{\bm{r}}]$ and bead velocities in the interval $[{\bm{V}},{\bm{V}}+d{\bm{V}}]$ at time $t$. The Fokker-Planck equation equivalent to (\ref{beadeqns1})-(\ref{beadeqns2}) may now be written down as
\begin{equation}
\frac{\partial\Psi}{\partial{t}} + \nabla_{{\bm{r}}}\cdot({\bm{V}}\Psi) + \frac{1}{m}\nabla_{{\bm{V}}}\cdot((-\zeta({\bm{V}}-{\bm{v}})+{\bm{F}})\Psi) = \frac{\zeta{k_B}T}{m^2}\nabla_{{\bm{V}}}^2\Psi.\label{eq:FPfull1}
\end{equation}

Equation (\ref{eq:FPfull1}) should be equipped with some boundary conditions. We assume that the velocities $\bm{V}_i$ can take any value in $\mathbb{R}^d$ and $\Psi$ vanishes sufficiently fast when $\bm{V}_i\to\infty$.  We do not specify any particular boundary conditions for $\Psi$ as  $\bm{r}_i$ approaches $\partial\Omega$ for the moment. However, if there is no flux of the fluid across the boundary of $\Omega$, these boundary conditions should be chosen so that
\begin{align}\label{boundVN}
\int\int_{\partial\Omega}\bm{V}_2\cdot\bm{n}\Psi\;d\sigma_{\bm{r}_2}d\bm{V}=0,  \quad \forall \bm{r}_1\in\Omega,\\
\int\int_{\partial\Omega}\bm{V}_1\cdot\bm{n}\Psi\;d\sigma_{\bm{r}_1}d\bm{V}=0,  \quad \forall \bm{r}_2\in\Omega,\nonumber
\end{align}
where $d\sigma_{\bm{r}_i}$ denotes an element of the surface as $\bm{r}_i$ runs over $\partial\Omega$, and the integral sign without a subscript (here and elsewhere) indicates the integration over the whole of $\mathbb{R}^d$ or  $\mathbb{R}^{2d}$. Indeed, take any volume $\omega$ inside $\Omega$ and consider the dumbbells whose bead number 1 lies in $\omega$. The beads numbered 2 in these dumbbells cannot leave the domain $\Omega$ so that their flux through the boundary of $\Omega$ should be 0, i.e.
$$
\int\int_{\omega}\int_{\partial\Omega}\bm{V}_2\cdot\bm{n}\Psi\;d\sigma_{\bm{r}_2}d\bm{r}_1d\bm{V}=0.
$$
Since this reasoning holds for any $\omega\subset\Omega$, the first equation in  (\ref{boundVN}) should be satisfied for all $\bm{r}_1\in\Omega$. The second equation in  (\ref{boundVN}) is established in the same way.

We define the number density of dumbbells as
\begin{equation}
N(\bm{x},t):=\frac{1}{2}\left(\int\int_\Omega \Psi|_{\bm{r}_1=\bm{x}}\;d\bm{r}_2d\bm{V} + \int\int_\Omega \Psi|_{\bm{r}_2=\bm{x}}\;d\bm{r}_1d\bm{V}\right).
\label{eq:genNdef}
\end{equation}
and the average polymeric velocity as
\begin{equation}
\bm{v}_p:=\frac{1}{2N}\left(\int\int_\Omega \bm{V}_1 \Psi|_{\bm{r}_1=\bm{x}}\;d\bm{r}_2 d\bm{V} + \int\int_\Omega \bm{V}_2 \Psi|_{\bm{r}_2=\bm{x}}\;d\bm{r}_1 d\bm{V}\right).\label{eq:vpdef}
\end{equation}
The polymer volume fraction $\varphi$ is related to the number density $N$ through
\begin{equation}\label{fNV}
\varphi = NV_d,
\end{equation}
where $V_d$ denotes the volume of a single dumbbell.

It follows from the Fokker-Planck equation (\ref{eq:FPfull1}) that the number density $N$ should satisfy the continuity equation
\begin{equation}
\frac{\partial N}{\partial{t}} + \nabla_{\bm{x}}\cdot (\bm{v}_pN)=0.
\label{eqNt}
\end{equation}
To see this, let us integrate (\ref{eq:FPfull1}) with respect to
$\bm{r}_i$ over $\Omega$ to get, for $i=1,2$, respectively,
\begin{align}
\int\int_{\Omega}\frac{\partial\Psi}{\partial{t}}\;d\bm{r}_1d\bm{V} + \nabla_{\bm{r}_2}\cdot\int\int_{\Omega}
\bm{V}_2\Psi\;d\bm{r}_1d\bm{V} = 0,\label{eq:intFPr1}\\
\int\int_{\Omega}\frac{\partial\Psi}{\partial{t}}\;d\bm{r}_2d\bm{V} + \nabla_{\bm{r}_1}\cdot\int\int_{\Omega}
\bm{V}_1\Psi\;d\bm{r}_2d\bm{V} = 0, \label{eq:intFPr2}
\end{align}
where the divergence theorem in combination with (\ref{boundVN}) has been used for simplification.
Eqn. (\ref{eq:intFPr1}) is valid for any value of $\bm{r}_2$ and Eqn. (\ref{eq:intFPr2}) for any value of $\bm{r}_1$. We may therefore choose the value of $\bm{r}_2$ in (\ref{eq:intFPr1}) and of $\bm{r}_1$ in (\ref{eq:intFPr2}) to be equal to $\bm{x}$, add the two equations and divide by two. This gives us
\[
\frac{\partial N}{\partial{t}} + \frac{1}{2}\nabla_{\bm{x}}\cdot
\left(\int\int_\Omega \bm{V}_2 \Psi|_{\bm{r}_2=\bm{x}}\;d\bm{r}_1 d\bm{V} + \int\int_\Omega \bm{V}_1 \Psi|_{\bm{r}_1=\bm{x}}\;d\bm{r}_2 d\bm{V} \right)=0,
\]
which is the same as (\ref{eqNt}).
\subsection{Continuity equations}
The continuity equations for the two fluid phases may be expressed as
\begin{align}
\frac{\partial\varphi}{\partial{t}} + \nabla_{\bm{x}}\cdot(\varphi \bm{v}_p) & =0,\label{eq:vp}\\
\frac{\partial(1-\varphi)}{\partial{t}} + \nabla_{\bm{x}}\cdot((1-\varphi) \bm{v}_s) &=0\label{eq:vs}.
\end{align}
Note that despite the similarity between (\ref{eq:vp}) and (\ref{eq:vs}) their mathematical role in our modelling is different: the polymer phase continuity equation (\ref{eq:vp}) follows immediately from (\ref{fNV}) and (\ref{eqNt}) while the solvent continuity equation (\ref{eq:vs}) is independent and expresses the incompressibility of the solvent. Eqn. (\ref{eq:vs}) just means that the volume of solvent in any small material volume $\delta{V(t)}$, containing at all times the same solvent molecules, remains unchanged even though the solvent fraction  $1-\varphi$ in the same material element may vary with time. The changes in $\varphi$ are due to the fact that as $\delta{V(t)}$ is tracked in time, polymer is continuously streaming through the surface of $\delta{V(t)}$ (because, in general, $\bm{v}_p \not=\bm{v}_s$). It is in the sense that the volume occupied by the same solvent molecules is time invariant that the solvent (and, indeed, the polymer) is incompressible.

The continuity equation for the volume-averaged velocity of the polymer-solvent solution ($\bm{u}$, say) may be computed from the volume fraction average
\begin{equation}
\bm{u} = \varphi\bm{v}_p + (1-\varphi)\bm{v}_s.\label{eq:avmixvel}
\end{equation}
Adding the two continuity equations (\ref{eq:vp})-(\ref{eq:vs}), we see that the two-phase fluid is incompressible:
\begin{equation}
\nabla_{\bm{x}}\cdot\bm{u} = 0.\label{eq:divuiszero}
\end{equation}
\subsection{Linear momentum equations}
We turn now to the elaboration of the linear momentum equations for the solvent and polymer phases. Considering a material volume $\delta{V(t)}$ of a size greater than the size of a bead but smaller than the distance between the two ends of a dumbbell, we recognise
\begin{equation}
\frac{d}{dt}\int_{\delta{V(t)}}\rho_s(1-\varphi)\bm{v}_s\;dV,\label{eq:linmomsolv}
\end{equation}
as the rate of change of momentum of the solvent phase in $\delta{V(t)}$, where $\rho_s$ denotes the (constant) solvent density. Equating (\ref{eq:linmomsolv}) with the sum of forces acting on the solvent in $\delta{V(t)}$ and use of the Reynolds transport theorem and the solvent continuity equation (\ref{eq:vs}) yields, in the usual way (since $\delta{V(t)}$ is arbitrarily chosen) that
\begin{equation}
\rho_s(1-\varphi)\frac{D\bm{v}_s}{D t} = \eta_s\nabla_{\bm{x}}^2\bm{v}_s + \eta_s\nabla_{\bm{x}}(\nabla_{\bm{x}}\cdot\bm{v}_s)- \nabla_{\bm{x}}p_s + \bm{f},\label{eq:eqofmN1}
\end{equation}
where the first two terms term on the right-hand side of (\ref{eq:eqofmN1}) are due to viscous forces, $p_s$ is the pressure in the solvent and $\bm{f}=\bm{f}(\bm{x},t)$ is the density of the force exerted at a point $\bm{x}$ and at time $t$ on the solvent by the dumbbells. The material derivative here is associated with the solvent velocity, i.e. $\frac{D}{Dt}=\frac{\partial}{\partial t} + \bm{v}_s\cdot\nabla_{\bm{x}}$. The key observation in defining $\bm{f}$ is that all the interaction forces between the polymers and the solvent are contained in the equation of motion of dumbbells (\ref{beadeqns1}). Leaving to one side the spring force there (which acts inside the polymer phase), the action of the solvent on the polymer is described by the friction force $-\zeta(\bm{V}_i - \bm{v}_i)$ (on the bead number $i$) and the Brownian force. We argue furthermore that the Brownian forces do not contribute to $\bm{f}$. Indeed, they are a part of the mesoscopic  description of the fluid and they account for random kicks experienced by the polymer molecules from the surrounding solvent molecules. The quantity $\bm{f}$ on the other hand, exists on the macroscopic level, hence its computation involves averaging on volumes that are sufficiently large compared to the micro-scale. The random kicks should cancel each other during this averaging since the probability of any such kick is the same as that of one in the opposite direction. This observation is also consistent with the mathematical fact that the expectation of an increment of the Wiener process during any time interval is 0. Therefore, only the friction (hydrodynamic) forces should be incorporated into the formula for $\bm{f}$. By the principle of reciprocal action we thus obtain the following expression:
\begin{equation}
\bm{f}(\bm{x},t) = \int\int_{\Omega}\zeta(\bm{V}_1-\bm{v}_1)\;\Psi|_{\bm{r}_1=\bm{x}}\;d\bm{r}_2 d\bm{V}
   + \int\int_{\Omega}\zeta(\bm{V}_2-\bm{v}_2)\;\Psi|_{\bm{r}_2=\bm{x}}\;d\bm{r}_1 d\bm{V},\label{eq:eqofmp}
\end{equation}
where the first integral accounts for the dumbbells whose bead number 1 is at position $\bm{x}$ and the second integral accounts for those
dumbbells whose bead number 2 has position vector $\bm{x}$. By comparing (\ref{eq:eqofmp}) with the definitions (\ref{eq:genNdef}) and (\ref{eq:vpdef}) of the number density and the polymeric velocity and noting that $\bm{v}_i=\bm{v}_s(\bm{x})$ when $\bm{r}_i=\bm{x}$ we easily obtain another useful formula for the force $\bm{f}$:
\begin{equation}\label{f2nz}
\bm{f}=2N\zeta(\bm{v}_p-\bm{v}_s).
\end{equation}

The description of our model of the solution of dumbbells that takes into account the inertia of the beads is now complete. The system of governing equations thus consists of the Fokker-Planck equation
(\ref{eq:FPfull1}) for the polymer phase and the Navier-Stokes-type equations  (\ref{eq:vs})--(\ref{eq:eqofmN1}) for the solvent phase coupled through the expressions for the number density, volume fraction and inter-phase force density, i.e. (\ref{eq:genNdef}), (\ref{fNV}) and (\ref{eq:eqofmp}), respectively. We emphasize that these equations do not involve the averaged polymer velocity $\bm{v}_p$ which, together with the total velocity $\bm{u}$, should be regarded in our framework as a quantity emerging in a post-processing of the basic equations. Indeed, $\bm{v}_p$ can be always computed via (\ref{eq:vpdef}) once the distribution function $\Psi$ is known. In particular, we don't really need a linear momentum equation for the polymer phase. However, it is interesting to write down such an equation in order to compare it with the corresponding equation for the solvent phase (\ref{eq:eqofmN1}) and to see the relations of our modelling to the previous work of other authors. Assuming that the fluid fills the whole space or that $\Psi$ vanishes on $\partial\Omega$, this equation reads (see the derivation in Appendix B)
\begin{equation}
\rho_p\varphi
\left(\frac{\partial\bm{v}_p}{\partial t}+\bm{v}_p\cdot\nabla_{\bm{x}}\bm{v}_p\right)
 = -2N\zeta(\bm{v}_p-\bm{v}_s) + \nabla_{\bm{x}}\cdot\tilde{\bm{\tau}}^s - \rho_p\nabla_{\bm{x}}\cdot(\varphi\text{Var}(\bm{V})) ,\label{eq:eqofmp2}
\end{equation}
where $\rho_p:= 2m/V_d$ is the polymer density, and $\text{Var}(\bm{V})$ is the variance of $\bm{V}$ defined by
\begin{align}
\text{Var}(\bm{V})=\frac{1}{2N}\Big( &
\int\int_\Omega (\bm{V}_1-\bm{v}_p)(\bm{V}_1-\bm{v}_p) \Psi|_{\bm{r}_1=\bm{x}}\;d\bm{r}_2 d\bm{V} \notag\\
&+ \int\int_\Omega (\bm{V}_2-\bm{v}_p)(\bm{V}_2-\bm{v}_p) \Psi|_{\bm{r}_2=\bm{x}}\;d\bm{r}_1 d\bm{V}\Big).
\label{varvp}
\end{align}
The tensor $\tilde{\bm{\tau}}^s$ in (\ref{eq:eqofmp2}) is defined as
\begin{equation}
\tilde{\bm{\tau}}^s(\bm{x},t) = \int_{-1/2}^{1/2}\int\int \Psi|_{\bm{r}_1=\bm{x}+(s-\frac 12)\bm{q} \atop \bm{r}_2=\bm{x}+(s+\frac 12)\bm{q}}
\bm{q}\bm{F}(\bm{q})\;d\bm{V}d\bm{q}ds
\label{deftauss}
\end{equation}
This is a modification of the Kramers expression for the polymeric contribution to the stress tensor, in which only the inter-bead spring force is taken into account. A similar tensor arises ${\bm{\tau}}^s$ will arise in our study of the vanishing bead inertia limit, cf. (\ref{deftau}) and Appendix A.  The conceptual difference between the linear momentum equation for the solvent (\ref{eq:eqofmN1}) and that for the polymer (\ref{eq:eqofmp}) resides in the presence of $\text{Var}(\bm{V})$ in the latter, which measures the fluctuations of the polymer velocity, while the fluctuations of the solvent velocity are disregarded. This is consistent with our basic assumption that the solvent can be wholly modelled on the macroscopic level while keeping a mesoscopic description for the polymer, as explained in Remark 1 on p. \pageref{ftnote}. Momentum balance equations for the two phases in the two fluid models of Doi and Onuki \cite{DoiOnuki92} and of Milner \cite{Milner91}, analogous to those in (\ref{eq:eqofmN1}) and (\ref{eq:eqofmp2}), may be found in Equations (4.12)-(4.13) and (14) of these papers, respectively. Note that their description of both phases is macroscopic which manifests itself in the absence of the terms like $\text{Var}(\bm{V})$.

\section{Negligible bead inertia}\label{sec:iless}
\subsection{The Fokker-Planck equation}\label{ssec:FPe}
We now proceed to obtain the Fokker-Planck equation for the dumbbell distribution function satisfied in the limit of negligible inertial forces. To this end, and following Degond and Liu \cite{DegondLiu09}, we use the following scaling and change of variables:
\begin{equation}\label{scal}
m = \varepsilon^2, \;\bm{p} = \varepsilon\bm{V}.
\end{equation}
To motivate this scaling, we note that the energy equipartition theorem yields $\left< mV_i^2 \right>=3k_BT$ in thermal equilibrium so that, supposing that the velocity distribution is not far from the equilibrium one, we get that the characteristic value of the velocity should be of order $\sqrt{k_BT/m}$, so that $\bm{V}$ behaves like $1/\varepsilon$ as $\varepsilon\to 0$.

Writing out a regular perturbation series for $\Psi$
\begin{equation}
\Psi = \Psi_0 + \varepsilon \Psi_1 + \varepsilon^2 \Psi_2 + \ldots,\label{eq:perturbPsi}
\end{equation}
we are looing for a closed-form equation for $\Psi_0$ in which the contributions of order $\varepsilon$ and higher are neglected.%
\footnote{Treating $\varepsilon$ as a small parameter is strictly speaking meaningless since it is a dimensional quantity. However, by performing the standard and well-established non-dimensionalization of all the variables we can easily see that the non-dimensional counterpart of $\varepsilon$ is $\sqrt{\lambda_B / \lambda_H}$ where $\lambda_B=m/\zeta$ is the characteristic time scale for the velocity fluctuations and $\lambda_H=\zeta/4H$ is the characteristic relaxation time of the dumbbell, $H$ being the scale for the spring force. The condition $\lambda_B<<\lambda_H$ is fairly well satisfied in most experiments, cf. \cite{Schieber92, JSO88}. Here, we keep the dimensional parameter $\varepsilon$ only to make the notation more readable.}
After the change of variables (\ref{scal}), the Fokker-Planck equation (\ref{eq:FPfull1}) becomes
\begin{equation}
\frac{\partial\Psi}{\partial{t}} + \frac{1}{\varepsilon}\nabla_{\bm{r}}\cdot(\bm{p}\Psi) + \frac{1}{\varepsilon}\nabla_{\bm{p}}\cdot\left((\zeta{\bm{v}}+\bm{F})\Psi\right) = \frac{1}{\varepsilon^2}\zeta Q(\Psi),\label{eq:FPfull2}
\end{equation}
where
\begin{equation}
Q(\Psi):=\nabla_{\bm{p}}\cdot(\bm{p}\Psi) + k_BT\nabla_{\bm{p}}^2\Psi.\label{eq:Q1}
\end{equation}
Substituting the perturbation series for $\Psi$ (\ref{eq:perturbPsi}) into (\ref{eq:FPfull2}) leads to
\begin{equation}
\frac{\partial\Psi_{k-1}}{\partial{t}} + \nabla_{\bm{r}}\cdot(\bm{p}\Psi_k) + \nabla_{\bm{p}}\cdot\left((\zeta{\bm{v}}+\bm{F})\Psi_k\right) = \zeta Q(\Psi_{k+1}),\label{eq:FPfull3}
\end{equation}
for $k=-1,0,1,\ldots$, where we understand $\Psi_{-1}$ and $\Psi_{-2}$ to be zero. We introduce the marginal distribution functions $\psi_k(\bm{r},t)$ as
\begin{equation}
\psi_k(\bm{r},t) = \int \Psi_k\;d\bm{p}, \quad\quad (k=0,1,2,\ldots).\label{eq:defnpsi}
\end{equation}
Then, integrating (\ref{eq:FPfull3}) throughout with respect to $\bm{p}$ over all $\bm{p}- space$ we arrive at
\begin{equation}
\frac{\partial\psi_k}{\partial{t}} = - \nabla_{\bm{r}}\cdot\int\bm{p}\Psi_{k+1}\;d\bm{p},\quad\quad (k=0,1,2,\ldots).\label{eq:FPpsi}
\end{equation}

We first set $k=-1$ in (\ref{eq:FPfull3}) which yields $Q(\Psi_0)=0$. Now, from (\ref{eq:Q1}) it is not hard to see that $Q$ may be rewritten as
\begin{equation}
Q(\Psi) = k_BT \nabla_{\bm{p}}\cdot\left(M\nabla_{\bm{p}}\left(\frac{\Psi}{M}\right)\right),\label{eq:Q2}
\end{equation}
where $M$ is the Maxwellian function
\begin{equation}
M(\bm{p}) = C\exp(-p^2/2k_BT),\label{eq:defnMaxwell}
\end{equation}
$C$ is a normalizing constant (chosen to make the integral of $M$ equal to $1$) and $p^2:=\|\bm{p}\|_2^2$. From the divergence form of operator $Q$ in (\ref{eq:Q2}), it may be shown that the kernel of $Q$ is the linear span of the Maxwellian function $M(\bm{p})$ defined in (\ref{eq:defnMaxwell}), cf. \cite{DegondLiu09}. Consequently, $\Psi_0(\bm{r},\bm{p},t)$ for any fixed $\bm{r}$ and $t$ should be proportional to $M(\bm{p})$. In view of (\ref{eq:defnpsi}) and the normalization of $M(\bm{p})$, this means that $\Psi_0(\bm{r},\bm{p},t) = \psi_0(\bm{r},t) M(\bm{p})$.

The remainder of the task of obtaining a Fokker-Planck equation for $\psi_0$ will be occupied with deriving, when $k=0$, an expression for the right-hand side of (\ref{eq:FPpsi}) in terms of $\psi_0$. Note, therefore, that use of the identity
\begin{equation}
\bm{p} + \frac{k_BT}{M}\nabla_{\bm{p}}M = \bm{0},\label{eq:Maxid}
\end{equation}
and repeated application of integration by parts gives for any $\Psi$
\begin{align}
\int\bm{p}Q(\Psi)\;d\bm{p} & = k_BT\int \bm{p}\nabla_{\bm{p}}\cdot\left(M\nabla_{\bm{p}}\left(\frac{\Psi}{M}\right)\right)\;d\bm{p} = -k_BT\int M\nabla_{\bm{p}}\left(\frac{\Psi}{M}\right)\;d\bm{p},\notag\\
& = k_BT\int\frac{\Psi}{M}\nabla_{\bm{p}}M\;d\bm{p} = -\int\bm{p}\Psi\;d\bm{p}.\label{eq:pQ}
\end{align}
Eqn. (\ref{eq:FPfull3}) provides another way of deriving an expression for the first integral in (\ref{eq:pQ}) when $\Psi$ is replaced by $\Psi_{k+1}$: multiplying (\ref{eq:FPfull3}) throughout by $\bm{p}/\zeta$ and integrating with respect to $\bm{p}$ yields
\begin{align}
\int\bm{p}Q(\Psi_{k+1})\;d\bm{p} & = \frac{1}{\zeta}\left(\frac{\partial}{\partial{t}}\int \bm{p}\Psi_{k-1}\;d\bm{p} + \nabla_{\bm{r}}\cdot\int \bm{p}\bm{p}\Psi_k\;d\bm{p} \right. \nonumber\\
& \hspace{3cm} \left.+ \int \bm{p}\nabla_{\bm{p}}\cdot((\zeta{\bm{v}}+\bm{F})\Psi_k)\;d\bm{p}\right).\notag %\label{eq:FPfullint}\\
\end{align}
Performing an integration by parts on the last term here and combining it with (\ref{eq:pQ}) we arrive at
\begin{equation}
 -\int \bm{p}\Psi_{k+1}\;d\bm{p} = \frac{1}{\zeta}\left(\frac{\partial}{\partial{t}}\int \bm{p}\Psi_{k-1}\;d\bm{p} + \nabla_{\bm{r}}\cdot\int \bm{p}\bm{p}\Psi_k\;d\bm{p}-(\zeta{\bm{v}}+\bm{F})\psi_k\right).\label{eq:FPfulltimesp}
\end{equation}

Now set $k=0$ in (\ref{eq:FPfulltimesp}). As remarked already, $\Psi_{-1}\equiv 0$. Moreover, using the identity (\ref{eq:Maxid}) and integration by parts we
see that
\begin{align}
\int \bm{p}\bm{p}\Psi_0\;d\bm{p} & = \psi_0\int \bm{p}\bm{p}M\;d\bm{p} = -k_BT\psi_0\int \bm{p}\left(\nabla_{\bm{p}}M\right)\;d\bm{p}\nonumber\\
& = k_BT\psi_0\bm{\delta}\int M\;d\bm{p}=k_BT\psi_0 \bm{\delta}.\label{eq:ppMsimple}
\end{align}
Therefore,  (\ref{eq:FPfulltimesp}) with $k=0$ is simplified to
\begin{equation}
 \int \bm{p}\Psi_{1}\;d\bm{p} = \frac{1}{\zeta}\left((\zeta{\bm{v}}+\bm{F})\psi_0 - k_BT\nabla_{\bm{r}}\psi_0 \right).\label{pPsi1}
\end{equation}
From this, setting $k=0$ in (\ref{eq:FPpsi}) we get the closed equation for $\psi_0$ which is now a function of only $\bm{r}$ and $t$. This is the well-known Fokker-Planck equation for a dilute solution of dumbbells in the inertialess ($\varepsilon\rightarrow 0$) case:
\begin{equation}
\frac{\partial\psi_0}{\partial{t}} = \frac{1}{\zeta}\nabla_{\bm{r}}\cdot\left(k_BT\nabla_{\bm{r}}\psi_0 - (\zeta{\bm{v}}+\bm{F})\psi_0\right).\label{eq:psi0final}
\end{equation}
The right-hand side of (\ref{eq:psi0final}) is just a compact notation for
\[\frac{1}{\zeta}\sum_{i=1}^2\nabla_{\bm{r}_i}\cdot\left(k_BT\nabla_{\bm{r}_i}\psi_0-(\zeta{\bm{v}}_i+\bm{F}_i)\psi_0\right).\]

The Fokker-Planck equation may be recast into a more familiar form in terms of the position vector of the centre of mass, $\bm{x}$, and the
end-to-end vector, $\bm{q}$, defined as
\begin{equation}
\bm{x}:= \frac{1}{2}(\bm{r}_1+\bm{r}_2),\;\bm{q}:=\bm{r}_2-\bm{r}_1.\label{eq:xandq}
\end{equation}
For clarity of notation we will denote the function $\psi_0$ of $\bm{r}_1$, $\bm{r}_2$ and $t$ by $\psi$ when it is considered as a function of $\bm{x}$, $\bm{q}$ and $t$. We also recall the convention $\bm{F}_1=-\bm{F}_2=\bm{F}=\bm{F}(\bm{q})$.  Performing the change of variables from $\bm{r}_1,\ \bm{r}_2$ to $\bm{x},\ \bm{q}$ we thus arrive at
\begin{align}
\frac{\partial\psi}{\partial{t}} & = \nabla_{\bm{q}}\cdot\left(\frac{2k_BT}{\zeta}\nabla_{\bm{q}}\psi+\frac{2\bm{F}}{\zeta}\psi + (\bm{v}_1-\bm{v}_2)\psi\right)\nonumber\\
& \hspace{4cm} + \nabla_{\bm{x}}\cdot\left(\frac{k_BT}{2\zeta}\nabla_{\bm{x}}\psi - \left(\frac{\bm{v}_1+\bm{v}_2}{2}\right)\psi\right).\label{eq:psifinal}
\end{align}

\subsection{Polymer force acting on the solvent and the elastic stress calculator}\label{ssec:mbeK}
We now turn to the calculation of the force vector $\bm{f}$ appearing in (\ref{eq:eqofmN1}) in the case of negligible bead inertia.
By writing $\bm{V}_i = \bm{p}_i/\sqrt{m}$ ($=\bm{p}_i/\varepsilon$) and retaining only the $O(\varepsilon^0)$ terms in (\ref{eq:eqofmp}), the force vector is given by
\begin{align}
\bm{f}(\bm{x},t) &= \zeta\int\int_{\Omega}\bm{p}_1\Psi_1|_{\bm{r}_1=\bm{x}}\;d\bm{r}_2 d\bm{p} - \zeta\int_{\Omega}\bm{v}_1\psi_0|_{\bm{r}_1=\bm{x}}\;d\bm{r}_2 \nonumber\\
& + \zeta\int\int_{\Omega}\bm{p}_2\Psi_1|_{\bm{r}_2=\bm{x}}\;d\bm{r}_1 d\bm{p} - \zeta\int_{\Omega}\bm{v}_2\psi_0|_{\bm{r}_2=\bm{x}}\;d\bm{r}_1.\label{eq:expressionf}
\end{align}
The first moments of $\Psi_1$ can be related to $\psi_0$ as shown in equation (\ref{pPsi1}).
Substituting (\ref{pPsi1}) into (\ref{eq:expressionf}), we see that $\bm{f}$ can be rewritten as a sum of two components $\bm{f}^s$ (due to the spring) and $\bm{f}^t$ (due to thermal motion)
\begin{equation}
\bm{f}(\bm{x},t) = \bm{f}^s(\bm{x},t)+\bm{f}^t(\bm{x},t),\label{eq:f0}
\end{equation}
with
\begin{align}
\bm{f}^s(\bm{x},t) & = \int_{\Omega}\bm{F}_1\psi_0|_{\bm{r}_1=\bm{x}}\;d\bm{r}_2 + \int_{\Omega}\bm{F}_2\psi_0|_{\bm{r}_2=\bm{x}}\;d\bm{r}_1\label{fs},\\
\bm{f}^t(\bm{x},t) & =   - k_BT\left(
   \int_{\Omega}\nabla_{\bm{r}_1}\psi_0|_{\bm{r}_1=\bm{x}}\;d\bm{r}_2 + \int_{\Omega}\nabla_{\bm{r}_2}\psi_0|_{\bm{r}_2=\bm{x}}\;d\bm{r}_1\right).\label{ft}
\end{align}
In Appendix A it is shown that the component $\bm{f}^s$ can be represented as the divergence of a tensor $\bm{f}^s= \nabla_{\bm{x}}\cdot\bm{\tau}^s$ with
$\bm{\tau}^s$ given by
\begin{equation}\label{deftaus}
\bm{\tau}^s(\bm{x},t)
= \int_{-1/2}^{1/2}\int_{\mathcal{Q}_s(\bm{x})}\psi(\bm{x}+s\bm{q},\bm{q},t)\bm{q}\bm{F}\;d\bm{q}ds,
\end{equation}
where $\mathcal{Q}_s(\bm{x})$ is the set of vectors $\bm{q}$ such that $\bm{x}+(s\pm 1/2)\bm{q}\in\Omega$.
For the other component $\bm{f}^t$ of the force, we immediately obtain
\begin{equation}
\bm{f}^t(\bm{x},t)  = - k_BT\nabla_{\bm{x}}\left(
   \int_{\Omega}\psi_0|_{\bm{r}_1=\bm{x}}\;d\bm{r}_2 + \int_{\Omega}\psi_0|_{\bm{r}_2=\bm{x}}\;d\bm{r}_1\right)
   = - 2k_BT\nabla_{\bm{x}}N, \label{eq:dNdx}
\end{equation}
where we have used the formula for the dumbbell number density in the inertialess case
\begin{equation}\label{Ndef}
N(\bm{x},t)=\frac{1}{2}\left(
   \int_{\Omega}\psi_0|_{\bm{r}_1=\bm{x}}\;d\bm{r}_2 + \int_{\Omega}\psi_0|_{\bm{r}_2=\bm{x}}\;d\bm{r}_1\right),
\end{equation}
which is a direct consequence of the general definition  (\ref{eq:genNdef}).
Putting (\ref{deftaus}) and (\ref{eq:dNdx}) together we conclude that
\begin{equation}
\bm{f} = \nabla_{\bm{x}}\cdot\bm{\tau}^s- 2k_BT\nabla_{\bm{x}}N
= \nabla_{\bm{x}}\cdot\bm{\tau} - \nabla_{\bm{x}}p_p ,\label{eq:f0finalform}
\end{equation}
where, we have introduced the elastic stress tensor as
\begin{equation}
\bm{\tau}(\bm{x},t) := \int_{-1/2}^{1/2}\int_{\mathcal{Q}_s(\bm{x})}\psi(\bm{x}+s\bm{q},\bm{q},t)\bm{q}\bm{F}\;d\bm{q}ds - N(\bm{x},t)k_BT\bm{\delta}\label{deftau}
\end{equation}
and the polymeric contribution to the pressure as
\begin{equation}
p_p(\bm{x},t):=N(\bm{x},t)k_BT.\label{defpp}
\end{equation}
The reason for decomposing $\bm{\tau}^s$ into a sum of $\bm{\tau}$ and a pressure-like term is to ensure that the elastic stress $\bm{\tau}$ vanishes at equilibrium, consistent with classical definitions, as in \cite{Birdetal87}, for example.

Taking Eqns. (\ref{eq:eqofmN1}) and (\ref{eq:f0finalform}) together we may now write down the equation of linear momentum for the solvent phase as
\begin{equation}
\rho_s(1-\varphi)\frac{D\bm{v}_s}{D t} = \eta_s\nabla_{\bm{x}}^2\bm{v}_s + \eta_s\nabla_{\bm{x}}(\nabla_{\bm{x}}\cdot\bm{v}_s) - \nabla_{\bm{x}}p + \nabla_{\bm{x}}\cdot\bm{\tau},\label{eq:eqofmN3}
\end{equation}
where $p:=p_s+p_p$ is the total pressure.
\subsubsection*{Some remarks on (\ref{deftau})}
\begin{enumerate}
\item If we neglect the presence of the walls by writing $\mathcal{Q}_s(\bm{x}) = \mathcal{Q}$, independent of $\bm{x}$ and $s$, for any $\bm{x}$ and $s$, the subtraction of the isotropic term on the right-hand side of the expression
(\ref{deftau}) for the elastic stress tensor ensures that in equilibrium (i.e. when $\nabla\bm{v}=\bm{0}$) $\bm{\tau}\equiv\bm{0}$. See, for example Eqn. (13.2-18) of \cite{Birdetal87}.
\item In the non-homogeneous case, where full account is taken of wall effects, we note from (\ref{deftau}) that the elastic stress tensor $\bm{\tau}$ will not, in general, vanish in equilibrium. Moreover, on smooth boundaries $\partial\Omega$ which are impenetrable by the dumbbell beads the configuration space $\mathcal{Q}_s(\bm{x})$ has zero measure in $d$ dimensions, leading to the boundary evaluation of the elastic stress
\begin{equation}
\bm{\tau}(\bm{x},t) = -N(\bm{x},t) k_BT\bm{\delta} \quad \forall \bm{x}\in\partial\Omega.
\end{equation}
\item An expression  similar to (\ref{deftau}) for the elastic stress was derived by Biller and Petruccione in \cite{BillerP87, PetruccioneB88} and used in \cite{Lozinskietal04}.
\item We can provide an alternative approximate formula for the elastic stress that does not involve the auxiliary variable $s$. To this end, we recall the second line in  (\ref{fscalc}), which is a weak expression for $\bm{f}_s$ valid for any compactly supported test function $\bm{g}$, and transform it by expanding $\bm{g}(\bm{x}\pm\bm{q}/2)$ in a Taylor series about $\bm{x}$ and integrating by parts:
\begin{align}
\int_{\Omega}\bm{f}^s\cdot\bm{g}\;d\bm{x}
 &= -\int_{\Omega}\int_{\mathcal{Q}(\bm{x})}\bm{F}\psi\cdot\left(\bm{q}\cdot\nabla_{\bm{x}}\bm{g} + \frac{1}{24}(\bm{q}\cdot\nabla_{\bm{x}})^3\bm{g} +\cdots \right)\;d\bm{q}d\bm{x}\nonumber \\
 &=  \int_{\Omega} \bm{g}\cdot\nabla_{\bm{x}} \cdot \left( \int_{\mathcal{Q}(\bm{x})}\bm{q}\bm{F}\psi\;d\bm{q} \right)d\bm{x} \nonumber \\
  &\qquad   + \frac{1}{24}\int_{\Omega} \bm{g}\cdot\nabla_{\bm{x}} \cdot \left( \int_{\mathcal{Q}(\bm{x})}\bm{q}\bm{F}(\bm{q}\cdot\nabla_{\bm{x}})^2\psi\;d\bm{q} \right)d\bm{x}
  +\cdots
\nonumber
\end{align}
We thus see again that the vector $\bm{f}_s$ is represented as a divergence of a tensor field. Invoking the same expression for $\bm{f}_t$ 	as before and subtracting the isotropic term we obtain that the elastic stress introduced in (\ref{deftau}) can be also approximately computed via
\begin{align}
\bm{\tau}(\bm{x},t) = &\int_{\mathcal{Q}(\bm{x})}\bm{q}\bm{F} \psi(\bm{x},\bm{q},t)\;d\bm{q} + \frac{1}{24}\int_{\mathcal{Q}(\bm{x})}\bm{q}\bm{F}\left(\bm{q}\cdot\nabla_{\bm{x}}\right)^2\psi(\bm{x},\bm{q},t)\;d\bm{q}+\cdots\nonumber\\
& - N(\bm{x},t)k_BT\bm{\delta}.\label{eq:f0Taylor}
\end{align}
Retention of only the first term in the Taylor series in (\ref{eq:f0Taylor}) would then yield the usual Kramers expression for the elastic stress tensor:
\begin{equation}
\bm{\tau} =  \langle\;\bm{q}\bm{F}\;\rangle - N k_BT\bm{\delta},
\end{equation}
where we have used the notation $\langle\;\cdot\;\rangle$ for the ensemble average
\begin{equation}
\langle\;\cdot\;\rangle :=\int_\mathcal{Q}\;\cdot\;\psi(\bm{x},\bm{q},t)\;d\bm{q}.
\end{equation}
\end{enumerate}
\subsection{The polymer velocity and the number density equation}\label{ssec:Nde}
The polymer phase velocity, $\bm{v}_p$, defined by (\ref{eq:vpdef}), can be easily computed in the inertialess limit thanks to the formula (\ref{f2nz}). Indeed, substituting the expression for $\bm{f}$ from (\ref{eq:f0finalform}) we arrive at
\begin{equation}
\bm{v}_p = \bm{v}_s + \frac{1}{2N\zeta}\bm{f} = \bm{v}_s + \frac{1}{2\zeta{N}}\nabla_{\bm{x}}\cdot\bm{\tau} - \frac{k_BT}{2\zeta{N}}\nabla_{\bm{x}}N.\label{eq:vpformula}
\end{equation}

Inserting this expression for $\bm{v}_p$ into (\ref{eqNt}) we finally arrive at the equation for the polymer number density in the inertialess limit
\begin{equation}
\frac{\partial N}{\partial t}+  \nabla_{\bm{x}}\cdot(\bm{v}_sN) = \frac{k_BT}{2\zeta}\nabla_{\bm{x}}^2N - \frac{1}{2\zeta}\nabla_{\bm{x}}\nabla_{\bm{x}}:\bm{\tau},\label{eq:DNDteq}
\end{equation}
where we use the notation $\nabla_{\bm{x}}\nabla_{\bm{x}}:\bm{\tau}$ in (\ref{eq:DNDteq}) to mean the divergence of the divergence of $\bm{\tau}$, i.e.
\begin{equation}
\nabla_{\bm{x}}\cdot\left(\nabla_{\bm{x}}\cdot\bm{\tau}\right) = \sum_{i=1}^d\sum_{j=1}^d\frac{\partial^2\tau_{ij}}{\partial{x_i}\partial{x_j}}.
\end{equation}
Note that, unlike Beris and Mavrantzas \cite{BerisMav94}, we have not needed to use Taylor's theorem and the neglect of higher-order terms in
order to arrive at the convection-diffusion equation (\ref{eq:DNDteq}) for $N$. The exactness of this equation might explain, therefore, why it was that the consistent derivation of the non-homogeneous kinetic theory, the continuum two-fluid Hamiltonian model and the body-tensor continuum formalism considered by these authors resulted in identical descriptions of the equation satisfied by the polymer number density.

\subsection{Summary of the model with neglected bead inertia}\label{ssec:polymandsolvvels1}
 As the governing equations of the model in the inertialess case  are scattered throughout the article, it can be helpful to recapitalize them here as follows: the solvent velocity $\bm{v}_s$ is obtained from the equations of motion (\ref{eq:eqofmN3}) and (\ref{eq:vs}), where the elastic stress tensor $\bm{\tau}$ in (\ref{eq:eqofmN3}) is found from Eqns. (\ref{deftau})--(\ref{Ndef}) and $\psi$ in (\ref{deftau}) is the solution to the Fokker-Planck equation (\ref{eq:psifinal}) with $\bm{v}_i=\bm{v}_s(\bm{r}_i)$, while the polymer volume fraction $\varphi$ is calculated from $\psi$ through (\ref{Ndef}) and (\ref{fNV}). The continuity equation (\ref{eq:vs}) can be reaplaced by an equivalent equation for the divergence of $\bm{v}_s$:
\[\frac{1}{V_d}\nabla\cdot\bm{v}_s = \frac{k_BT}{2\zeta}\nabla^2N - \frac{1}{2\zeta}\nabla\nabla:\bm{\tau},\] which is easily derived with the aid of (\ref{eq:DNDteq}). If needed, the average polymer phase velocity $\bm{v}_p$ can be computed then by (\ref{eq:vpformula}) and the volume-averaged velocity of the solution $\bm{u}$ by (\ref{eq:avmixvel}).

\section{A modification of the theory with the assumption ``surrounding fluid = solution''}\label{sec:polymandsolvvels}
All of the preceding theory has been developed assuming ``surrounding fluid = solvent'' in the drag force term of (\ref{beadeqns1}). As has already been mentioned in Remark 1 on p. \pageref{ftnote}  this hypothesis is rather arbitrary although it seems appropriate in very dilute solutions and in the absence of hydrodynamic interaction. It is also the underlying assumption in the works of \cite{Doi90, DoiOnuki92, Milner91}. However, we may also consider an alternative hypothesis ``surrounding fluid = solution'' in (\ref{beadeqns1}), which can be seen as a very crude way to account for the hydrodynamic interaction.

The theory of Sections \ref{sec:KT}, \ref{sec:iless} may be easily modified to accommodate this case: in the dumbbell equations of motion (\ref{beadeqns1})-(\ref{beadeqns2}), for example, the only change is to set $\bm{v}_i=\bm{u}(\bm{r}_i)$, $i=1,2$, $\bm{u}$ being the averaged solution velocity. The kinetic theory for the polymer phase remains essentially the same and the Fokker-Planck equation (\ref{eq:psifinal}) still holds in the inertialess limit, again assuming $\bm{v}_i=\bm{u}(\bm{r}_i)$. However, some changes should be made to the coupling of these mesoscopic equations and the macroscopic motion equation of motion for the solvent (\ref{eq:eqofmN1}). In the equations below, and for ease of notation, let us keep the definition of the force $\bm{f}$ to be the friction force $2N\zeta(\bm{v}_p-\bm{v}_s)$ between the polymer and the solvent, as it was in the preceding section.

Without entering into all the details of the derivation, which are essentially the same as in the preceding section, we just summarize here the governing equations stemming from the hypothesis ``surrounding fluid = solution''. They include the Fokker-Planck equation (\ref{eq:psifinal}) with the substitution $\bm{v}_1=\bm{u}(\bm{x}-\bm{q}/2,t)$, $\bm{v}_2=\bm{u}(\bm{x}+\bm{q}/2,t)$, while the equations for the macroscopic quantities $\bm{v}_s$, $\bm{u}$ and $\bm{\tau}$ become
\begin{equation}
\rho_s(1-\varphi)\frac{D\bm{v}_s}{D t} = \eta_s\nabla_{\bm{x}}^2\bm{v}_s + \eta_s\nabla_{\bm{x}}(\nabla_{\bm{x}}\cdot\bm{v}_s) - \nabla_{\bm{x}}p_s + \bm{f},\label{eq:eqofmN1bis}
\end{equation}
\[\frac{\partial(1-\varphi)}{\partial{t}} + \nabla_{\bm{x}}\cdot((1-\varphi) \bm{v}_s) =0,\]
\[\bm{\tau}(\bm{x},t) = \int_{-1/2}^{1/2}\int_{\mathcal{Q}_s(\bm{x})}\psi(\bm{x}+s\bm{q},\bm{q},t)\bm{q}\bm{F}\;d\bm{q}ds - 2N(\bm{x},t)k_BT\bm{\delta},\]
\begin{equation}
\bm{u} = \bm{v}_s + \frac{V_d}{2\zeta(1-\varphi)}\nabla_{\bm{x}}\cdot\bm{\tau}, \label{eq:pertubvs}
\end{equation}
with $N$ calculated from (\ref{eq:DNDteq}) (being careful to replace $\bm{v}_s$ in this equation with $\bm{u}$). Since $(\bm{v}_p-\bm{u})=(1-\varphi)(\bm{v}_p-\bm{v}_s)$, the force $\bm{f}$ on the right-hand side of (\ref{eq:eqofmN1bis}) can now be calculated from
\[\bm{f} = \frac{1}{1-\varphi}\nabla_{\bm{x}}\cdot\bm{\tau}.\]
Note, therefore, that the force between the polymeric and solution phases in Eqn. (\ref{eq:eqofmN1bis}) is no longer the divergence of a tensor because of the division by $(1-\varphi)$. For the same reason, we are not easily able to interpret a part of the elastic stress as a pressure-like term.

By writing the solution velocity $\bm{u}$ as the sum of the solvent velocity $\bm{v}_s$ and a disturbance velocity ${\bm{v}^\prime_s}$, we see from (\ref{eq:f0finalform}) and (\ref{eq:pertubvs}) that
\begin{equation}
{\bm{v}^\prime_s} = \frac{V_d}{2\zeta}\bm{f} = \frac{\varphi}{\zeta(1-\varphi)}\zeta(\bm{v}_p - \bm{u}) = \frac{\varphi}{\zeta(1-\varphi)}\zeta(\bm{v}_p - (\bm{v}_s+{\bm{v}^\prime_s})).\label{eq:disturbvel}
\end{equation}
Although Eqn. (\ref{eq:disturbvel}) holds as a direct consequence of the definition (\ref{eq:avmixvel}) of the solution velocity $\bm{u}$ whether we consider the surrounding fluid to be the solvent or the solution, the interpretation is rather different in the two cases. By ``surrounding fluid'' we really mean the fluid in the immediate neighbourhood of a dumbbell. In Section \ref{ssec:polymandsolvvels1}, ${\bm{v}^\prime_s}$ would just be the difference between the solution velocity and the solvent velocity, although the velocity field in the immediate neighbourhood of any individual dumbbell is considered to be affected by other dumbbells only in so far as they contribute to the elastic stress in the equation of motion (\ref{eq:eqofmN3}). When the surrounding fluid is considered to be the solution, ${\bm{v}^\prime_s}$ is the actual disturbance to the velocity field in the immediate neighbourhood of a bead caused by the other dumbbells beyond their contribution to the elastic stress in the solvent equation of motion. From (\ref{eq:disturbvel}), ${\bm{v}^\prime_s}$ is seen to be the product of a function and the force exerted by the polymer phase on the solution and is much simpler than in the case of full hydrodynamic interaction.

\section{Non-dimensionalization of the governing equations and elaboration of important limiting cases}\label{ssec:elaborateeqns}
Let $V$ and $L$ denote characteristic velocity and macroscopic length scales, respectively, and $N_{av}$ be a (space- and time-) averaged value of $N$. The homogeneous version of the Fokker-Planck equation (\ref{eq:psifinal}) for a Hookean dumbbell in equilibrium ($\bm{v}_s=0$) tells us that a characteristic mesoscopic length scale $\ell_0:=\sqrt{{\rm{tr}}\langle\bm{qq}\rangle/d}$ satisfies $\ell_0^2H=k_BT$ where $H$ is the spring force constant for a Hookean dumbbell.

We may introduce non-dimensional variables:
\begin{equation}
\bm{x}^*=\frac{\bm{x}}{L},\;\bm{v}_s^*=\frac{\bm{v}_s}{V},\;t^*=\frac{tV}{L},\;N^* = \frac{N}{N_{av}},\;\bm{q}^*=\frac{\bm{q}}{\ell_0},\;\bm{F}^* = \frac{\bm{F}}{H\ell_0}.\label{eq:rescale}
\end{equation}
This rescaling leads to the non-dimensional Fokker-Planck equation, cf. (\ref{eq:psifinal}),
\begin{align}
\frac{D\psi^*}{D{t^*}} & = \nabla_{\bm{q}^*}\cdot\left(\frac{1}{2De}\nabla_{\bm{q}^*}\psi^* + \frac{\bm{F}^*}{2De}\psi^*-\left\{(\bm{q}^*\cdot\nabla_{\bm{x}^*})\bm{v}_s^* + \frac{1}{24}\left(\frac{\ell_0}{L}\right)^2(\bm{q}^*\cdot\nabla_{\bm{x}^*})^3\bm{v}_s^*+\ldots\right\}\psi^*\right)\nonumber\\
& + \nabla_{\bm{x}^*}\cdot\left(\frac{1}{8De}\left(\frac{\ell_0}{L}\right)^2\nabla_{\bm{x}^*}\psi^* - \left\{ \frac{1}{4}\left(\frac{\ell_0}{L}\right)^2(\bm{q}^*\cdot\nabla_{\bm{x}^*})^2\bm{v}_s^*+\cdots\right\}\psi^*\right),\label{eq:FPxqnd}
\end{align}
where
\begin{equation}
De:=\frac{\zeta{V}}{4HL},\label{eq:De}
\end{equation}
is a Deborah number and the non-dimensional distribution function $\psi^*$ is introduced so that $\psi=\psi^*N_{av}/\ell_0^d$. The terms discarded in (\ref{eq:FPxqnd}) are of order 4 and higher in $\ell_0/L$.

In order to give a reasonable scaling for the elastic stress we remark that it is of order $N_{av}k_BT\zeta/(4H)(\nabla\bm{v}+\nabla\bm{v}^T)$ in the regime of small deviations from equilibrium (in fact, this is the first term in the Taylor series for $\bm{\tau}$ as $\nabla\bm{v}\to 0$ in the case of Hookean dumbbells). The dimensionless elastic stress tensor $\bm{\tau}^*$ is thus introduced as
\begin{equation}
\bm{\tau}^*=\frac{\bm{\tau}}{N_{av}k_BTDe},\label{rescal_tau}
\end{equation}
and the expression (\ref{eq:f0Taylor}) for $\bm{\tau}$ becomes
\begin{align}
\bm{\tau}^*(\bm{x}^*,t^*)  = \frac{1}{De}\Big[
& \int_{\mathcal{Q}^*(\bm{x}^*)}  \bm{q}^* \bm{F}^* \psi^*(\bm{x}^*,\bm{q}^*,t)\;d\bm{q}^* \nonumber\\
& +  \frac{1}{24}\left(\frac{\ell_0}{L}\right)^2  \int_{\mathcal{Q}^*(\bm{x}^*)}  \bm{q}^* \bm{F}^* (\bm{q}^*\cdot\nabla_{\bm{x}^*})^2 \psi^*(\bm{x}^*,\bm{q}^*,t)\;d\bm{q}^*
    + \ldots \nonumber\\
& - N^*\bm{\delta}\Big],\label{eq:nondimdivT}
\end{align}
where we have written out only those terms up to $O(\ell_0/L)^2$, as in (\ref{eq:FPxqnd}). The expression (\ref{eq:vpformula}) for the polymer phase velocity becomes
\begin{equation}
\bm{v}_p^*=\bm{v}_s^*
+ \frac{1}{8N^*}\left(\frac{\ell_0}{L}\right)^2 \left( \nabla_{\bm{x}^*}\cdot\bm{\tau}^* - \frac{1}{De}\nabla_{\bm{x}^*}N^* \right),\label{eq:vpnondim}
\end{equation}The dimensionless number density equation (\ref{eq:DNDteq}) now reads
\begin{equation}
\frac{\partial N^*}{\partial t^*} +  \nabla_{\bm{x}^*}\cdot(\bm{v}_s^*N^*)
= \frac{1}{8}\left(\frac{\ell_0}{L}\right)^2 \left(\frac{1}{De}\nabla_{\bm{x}^*}^2N^* - \nabla_{\bm{x}^*}\nabla_{\bm{x}^*}:\bm{\tau}^* \right).\label{eq:DNDtnondim}
\end{equation}
Finally, the characteristic total viscosity may be introduced as $\eta=\eta_s+N_{av}k_BT\zeta/(4H)$ and the corresponding Reynolds number is $Re=\rho_sVL/\eta$. This leads to the non-dimensional equation of motion
\begin{align}
Re(1-\varphi)\frac{D\bm{v}_s^*}{Dt^*} &= -\nabla_{\bm{x}^*}p^* + \eta^*_s\nabla_{\bm{x}^*}^2\bm{v}_s^* + \eta^*_p\nabla_{\bm{x}^*}\cdot\bm{\tau}^*,\label{eq:dimlesseom}
\end{align}
where $\eta^*_s$ and $\eta^*_s$ are dimensionless solvent and polymer viscosities defined, respectively, as $\eta^*_s=\eta^*_s/\eta$ and $\eta^*_p=N_{av}k_BT\zeta/(4H\eta)$.

\subsection{$\ell_0/L$ negligible}\label{ssec:l0neg}
Let us first consider the equation set to be solved in the case of $\ell_0/L$ negligible (known as locally homogeneous flow).
In this case, there is no distinction between the phase velocities $\bm{v}_p^*=\bm{v}_s^*=\bm{u}^*$ and the Fokker-Planck equation (\ref{eq:FPxqnd}) assumes its standard form, in particular there is no diffusion in $\bm{x}$ there.  Moreover, as remarked already in Section \ref{ssec:mbeK} and as may be seen from (\ref{eq:nondimdivT}), the Kramers expression holds for $\bm{\tau}^*$
\begin{equation}
\bm{\tau}^* = \langle \bm{q}^*\bm{F}^* \rangle^* - N^*\bm{\delta},\label{eq:Kramershom}
\end{equation}
where the average is taken over the configuration space $\mathcal{Q}^*$, independent of $\bm{x}^*$. As is well known, a closed-form differential equation for $\langle\bm{q}^*\bm{F}^*\rangle^*$
may be obtained in the case of FENE-P or Hookean dumbbells, for example, by multiplying the Fokker-Planck equation (\ref{eq:FPxqnd}) throughout by $\bm{q}^*\bm{F}^*$ and using integration by parts. This equation, together with the equations of motion (\ref{eq:dimlesseom}) for  $\bm{u}^*=\bm{v}_s^*$, the incompressibility equation $\nabla\cdot\bm{u}^*=0$, and the number density equation
$DN^*/Dt^* = 0$,
now form a solvable system, to be solved subject to suitable initial and inflow conditions.

\subsection{Non-homogeneous flows}
The major impediment to elaborating a system of equations to be solved in the non-homogeneous case, even when the dumbbell spring is Hookean, is that
retention of terms of $O(\ell_0^2/L^2)$ in (\ref{eq:FPxqnd}), (\ref{eq:nondimdivT}) and (\ref{eq:DNDtnondim}) means that the system cannot be closed even if one persuades oneself that the $O(\ell_0^2/L^2)$ terms in the Fokker-Planck equation (\ref{eq:FPxqnd}) may be neglected when the length scale for velocity variations is comparable to $\ell_0$. See the arguments used by Beris and Mavrantzas \cite{BerisMav94}, for example. This is because, in contradiction to the form of the Kramers expression (\ref{eq:Kramershom}) assumed by Bhave et al. \cite{Bhaveetal91} and Beris and Mavrantzas \cite{BerisMav94}, for example, one must, for the sake of consistency with what is done elsewhere, retain all terms up to $O(\ell_0^2/L^2)$ in the expression (\ref{eq:nondimdivT}) for $\bm{\tau}$. On the other hand, if we are prepared to sacrifice consistency and to keep the terms of order $(\ell_0/L)^2$ only in the $\bm{x}$-diffusion term in (\ref{eq:FPxqnd}) together with (\ref{eq:vpnondim}) and (\ref{eq:DNDtnondim}), a closed-form constitutive equation for $\bm{\tau}$ can be derived in our framework as well.

\section*{Conclusions}
We have presented a framework to model the dilute polymer solution in the spirit of the two fluid theory. Unfortunately, the obtained equations do not reduce to simpler constitutive equations in the case of strongly non-homogeneous flows in any self-consistent way, even if Hookean dumbbells are used to represent the polymer molecules. One should hope, however, that their numerical simulation is possible by stochastic or Fokker-Planck-based methods as is demonstrated, for example, in \cite{Lozinskietal04} for a similar model. We believe that the use of a perturbation series in going from the equations of motion with bead inertia to the inertialess limit provides a pleasing and rigorous method of deriving both the Fokker-Planck equation and a generalized Kramers-type expression for the elastic stress for this case. Moreover, this approach may be pursued further in keeping more terms in the perturbation series. This will provide not only additional terms in the momentum balance equation but also in that of the energy equation.

\appendix
\section{The averaged spring force in the inertialess case}
Multiplying the formula (\ref{fs}) for the averaged spring force $\bm{f}^s$ by an arbitrary vector-valued test function $\bm{g}$ vanishing on the boundary $\partial\Omega$, integrating over $\Omega$ and changing the variables in the integral from $d\bm{r}_1d\bm{r}_2$ to $d\bm{x}d\bm{q}$, we see that
\begin{align}
\int_{\Omega}\bm{f}^s(\bm{x},t)\cdot\bm{g}(\bm{x})\;d\bm{x}&=
\int_{\Omega}\int_{\Omega}\bm{F}_1\psi_0\cdot\bm{g}(\bm{r}_1)\;d\bm{r}_1d\bm{r}_2 + \int_{\Omega}\int_{\Omega}\bm{F}_2\psi_0\cdot\bm{g}(\bm{r}_2)\;d\bm{r}_1d\bm{r}_2 \nonumber\\
& = \int_{\Omega}\int_{\mathcal{Q}(\bm{x})}\bm{F}\psi\cdot\left(\bm{g}(\bm{x}-\bm{q}/2)-\bm{g}(\bm{x}+\bm{q}/2)\right)\;d\bm{q}d\bm{x}\label{fscalc} \\
& = - \int_{\Omega}\int_{\mathcal{Q}(\bm{x})}\bm{F}\psi(\bm{x},\bm{q},t)\cdot\int_{-1/2}^{1/2}\bm{q}\cdot\nabla\bm{g}(\bm{x}-s\bm{q})\;dsd\bm{q}d\bm{x},
\nonumber
\end{align}
where $\mathcal{Q}(\bm{x})$ is the set of vectors $\bm{q}$ such that $\bm{x}\pm\bm{q}/2\in\Omega$. By making the change of variable $\bm{x}\to\bm{x}+s\bm{q}$, we arrive at
\[
\int_{\Omega}\bm{f}^s(\bm{x},t)\cdot\bm{g}(\bm{x})\;d\bm{x}
= - \int_{\Omega}\int_{-1/2}^{1/2}\int_{\mathcal{Q}_s(\bm{x})}\psi(\bm{x}+s\bm{q},\bm{q},t)\bm{F}\cdot(\bm{q}\cdot\nabla\bm{g})(\bm{x})\;d\bm{q}dsd\bm{x},
\]
where $\mathcal{Q}_s(\bm{x})$ is the set of vectors $\bm{q}$ such that $\bm{x}+(s\pm 1/2)\bm{q}\in\Omega$. Using integration by parts, this leads to
\begin{equation}\label{eq:2Fterms}
\bm{f}^s(\bm{x},t)
= \nabla_{\bm{x}}\cdot \int_{-1/2}^{1/2}\int_{\mathcal{Q}_s(\bm{x})}\psi(\bm{x}+s\bm{q},\bm{q},t)\bm{q}\bm{F}\;d\bm{q}ds.
\end{equation}
We recognize that the right-hand side here is the divergence of tensor $\bm{\tau}^s$ as defined in (\ref{deftaus}).
\section{Derivation of the averaged momentum equation for the polymer phase (%
\protect\ref{eq:eqofmp2})}

Let us first note that the definitions of the number density (\ref%
{eq:genNdef}), the average polymeric velocity (\ref{eq:vpdef}) and its
variance (\ref{varvp}) can be simplified thanks to the symmetry $\Psi (%
\bm{r}_{1},\bm{r}_{2},\bm{V}_{1},\bm{V}_{2},t)=\Psi (\bm{r}_{2},\bm{r}_{1},\bm{V}_{2},\bm{V}_{1},t)$:
\begin{equation*}
N(\bm{x},t)=\int \int_{\Omega }\Psi |_{\bm{r}_{1}=\bm{x}}\;d%
\bm{r}_{2}d\bm{V},\quad \bm{v}_{p}(\bm{x},t)=\frac{1}{N}\int
\int_{\Omega }\bm{V}_{1}\Psi |_{\bm{r}_{1}=\bm{x}}\;d\bm{r}%
_{2}d\bm{V},
\end{equation*}%
\begin{equation*}
\text{Var}(\bm{V})=\frac{1}{N}\int \int_{\Omega }(\bm{V}_{1}-\bm{%
v}_{p})(\bm{V}_{1}-\bm{v}_{p})\Psi |_{\bm{r}_{1}=\bm{x}}\;d%
\bm{r}_{2}d\bm{V.}
\end{equation*}%
This leads immediately to the following relation
\begin{equation}
\text{Var}(\bm{V})=\frac{1}{N}\int \int_{\Omega }\bm{V}_{1}\bm{V}%
_{1}\Psi |_{\bm{r}_{1}=\bm{x}}\;d\bm{r}_{2}d\bm{V-\bm{v}%
_{p}\bm{v}_{p}}.  \label{var1}
\end{equation}
Let us now proceed to the derivation of (\ref{eq:eqofmp2}) starting from the
Fokker-Planck equation (\ref{eq:FPfull1}). We multiply (\ref{eq:FPfull1}) by
$\bm{V}_{1}$ and integrate throughout with respect to $\bm{r}_{2}$
and $\bm{V}$. In doing this we denote $\bm{r}_{1}$ as $\bm{x}$
to be consistent with the formulas for $N$ and $\bm{v}_{p}$ above. After
an integration by parts with respect to $\bm{V}$, we thus obtain%
\begin{align*}
\frac{\partial }{\partial {t}}\int \int \bm{V}_{1}\Psi|_{\bm{r}_{1}=\bm{x}} d\bm{r}_{2}d%
\bm{V}
&+\nabla _{{\bm{x}}}\cdot \int \int {\bm{V}}_{1}{\bm{V}}%
_{1}\Psi|_{\bm{r}_{1}=\bm{x}} d\bm{r}_{2}d\bm{V}+\int \int {\bm{V}}_{1}{\bm{V}}%
_{2}\cdot \nabla _{{\bm{r}}_{2}}\Psi|_{\bm{r}_{1}=\bm{x}} d\bm{r}_{2}d\bm{V} \\
&=\frac{1}{m}\int \int (-\zeta ({\bm{V}}_{1}-{\bm{v}}_{1})+{\bm{F}}%
_{1})\Psi|_{\bm{r}_{1}=\bm{x}} d\bm{r}_{2}d\bm{V}.
\end{align*}%
The third term on the left-hand side can be reduced to an integral over $%
\partial \Omega $ with respect to $\bm{r}_{2}$ and thus it vanishes if $\Psi $ is zero on the boundary $\partial \Omega $ or, in particular, when $\Omega $ is the whole space $\mathbb{R}^{d}$. All the other terms in the
last equation are simplified with the aid of definitions of $N$ and $\bm{v}_{p}$ and relation (\ref{var1}):
\begin{equation}
\frac{\partial }{\partial {t}}(N\bm{v}_{p})+\nabla _{{\bm{x}}}\cdot
(N\bm{v}_{p}\bm{v}_{p}+N\text{Var}(\bm{V}))=-\frac{N\zeta }{m}({%
\bm{v}}_{p}-{\bm{v}}_{s})+\frac{1}{2m}{\bm{\tilde{f}}}^{s},
\label{mfp1}
\end{equation}
where $\bm{\tilde{f}}^{s}$ is the averaged spring force
\begin{align*}
\bm{\tilde{f}}^{s}(\bm{x},t)
&=\int \int \bm{F}_{1}\Psi |_{\bm{r}_{1}=\bm{x}}\;d\bm{r}_{2}d\bm{V}+\int \int \bm{F}_{2}\Psi|_{\bm{r}_{2}=\bm{x}}\;d\bm{r}_{1}d\bm{V} \\
&=2\int \int \bm{F}_{1}\Psi |
_{\bm{r}_{1}=\bm{x}}\;d\bm{r}_{2}d\bm{V}.
\end{align*}%
The last quantity is almost the same as $\bm{f}^{s}$ introduced by
formula (\ref{fs}) in the study of the inertialess case. The only difference
is that $\psi _{0}$ featuring in the definition of $\bm{f}^{s}$ is now
replaced by $\tilde{\psi}_{0}=\int \Psi d\bm{V}$. We can thus repeat the
reasoning of the preceding appendix B in order to rewrite $\bm{\tilde{f}}%
^{s}$ as the divergence of a tensor%
\begin{equation*}
\bm{\tilde{f}}^{s}=\nabla _{\bm{x}}\cdot \int_{-1/2}^{1/2}\int
\tilde{\psi}(\bm{x}+s\bm{q},\bm{q},t)\bm{q}\bm{F}\;d%
\bm{q}ds:=\nabla _{\bm{x}}\cdot \tilde{\bm{\tau}}^{s},
\end{equation*}%
where $\tilde{\psi}$ is $\ \tilde{\psi}_{0}$ rewritten in terms of $\bm{r%
}_{c}$ and $\bm{q}$. The definition of $\tilde{\bm{\tau}}^{s}$ is rewritten in (\ref{deftauss}) in terms of $\Psi$. Multiplying (\ref{mfp1}) by $2m$, introducing the
polymer density $\rho _{p}=2m/V_{d}$ and recalling that $N=\varphi /V_{d}$,
we now obtain%
\begin{equation}
\frac{\partial }{\partial {t}}(\rho _{p}\varphi \bm{v}_{p})+\nabla _{{%
\bm{x}}}\cdot (\rho _{p}\varphi \bm{v}_{p}\bm{v}_{p})=-2N\zeta ({%
\bm{v}}_{p}-{\bm{v}}_{s})+\nabla _{\bm{x}}\cdot \tilde{\bm{\tau}}%
^{s}-\nabla _{{\bm{x}}}\cdot (\rho _{p}\varphi \text{Var}(\bm{V})).
\label{mfp2}
\end{equation}%
Combining (\ref{mfp2}) with the continuity equation (\ref{eqNt}) leads to (\ref{eq:eqofmp2}).

\end{document}